\documentclass{elsart5p}
\usepackage{graphicx}
\usepackage{amsfonts}
\usepackage{amsmath}
\usepackage{amssymb}

\allowdisplaybreaks[1]

\begin{document}

\begin{frontmatter}

\title{Supercontinuum generation of ultrashort laser pulses in air \\
at different central wavelengths}

\author{Stefan Skupin},
\ead{stefan.skupin@cea.fr}
\author{Luc Berg\'e}
\ead{luc.berge@cea.fr}

\address{D\'epartement de Physique Th\'eorique et Appliqu\'ee,
CEA/DAM Ile de France, B.P. 12, 91680 Bruy\`eres-le-Ch\^atel, France}

\begin{abstract}
Supercontinuum generation by femtosecond filaments in air is investigated for different laser wavelengths ranging from ultraviolet to infrared. Particular attention is paid on the role of third-harmonic generation and temporal steepening effects, which enlarge the blue part of the spectrum. A unidirectional pulse propagation model and nonlinear evolution equations are numerically integrated and their results are compared. Apart from the choice of the central wavelength, we emphasize the importance of the saturation intensity reached by self-guided pulses, together with their temporal duration and
propagation length as key players acting on both supercontinuum generation of the pump wave and emergence of the third harmonics. Maximal broadening is
observed for large wavelengths and long filamentation ranges.
\end{abstract}


\end{frontmatter}

\section{Introduction}

\label{Introduction}

Third harmonic (TH) generation and supercontinuum (SC) emission are
two phenomena which have attracted broad interest in the past years
\cite{Alfano:SLS:89,Chin:jnopm:8:121,Backus:ol:21:665,Fedotov:oc:133:587,Marcus:josab:16:792}. An evident reason is their direct application in
atmospheric remote sensing measurements based on LIDAR (LIght Detection And
Ranging) femtosecond laser setups \cite{Kasparian:sc:301:61}. In this
context, spectral broadening originates from complex mechanisms that
drive the long-range propagation of ultrashort pulses, when they form narrow filaments in optically-transparent media.

The physics of isolated femtosecond filaments in air is nowadays rather
well understood (see, e.g., \cite{Berge:review} and references therein). It involves the competition between Kerr self-focusing and plasma defocusing, triggered whenever the input pulse power exceeds the critical
power for self-focusing $P_{\rm cr} \simeq \lambda_0^2/(2 \pi n_0
n_2)$. Here, $\lambda_0$ is the central laser wavelength, $n_0 = 1$
and $n_2$ are the linear and nonlinear refraction indices in air,
respectively. For high enough powers, multiple filaments nucleated 
after an early stage of modulational instability have also been widely investigated \cite{Berge:review,Bespalov:jetp:3:307,Mlejnek:prl:83:2938}. 
They produce spectral patterns mostly analogous to those generated by a single
filament, as filamentary cells emerge in phase from the background field and
possess the same phase link \cite{Chin:oc:210:329}. By comparing Terawatt (TW)
multifilamented beams with Gigawatt (GW) single filaments in air, this
property was again verified in 
the UV-visible region (230-500~nm), where femtosecond self-focusing pulses centered at 800 nm generically produce a tremendous plateau of wavelengths \cite{Akozbek:apb:77:177,Theberge:apb:80:221,Berge:pre:71:016602}.

This latter phenomenon has recently become a subject of
inspiration for several researchers. Two scenarios have been proposed for
justifying the build-up of new wavelengths in the UV-visible
range. On the one hand, temporal steepening phenomena undergone by the pump were shown to deeply modify the filament spectrum \cite{Akozbek:oc:191:353}.
Full chromatic dispersion included in the optical field wave number 
$k(\omega)$ affects both the diffraction operator and the nonlinearities. This induces shock-like dynamics at the back edge of the pulse through space-time focusing and self-steepening effects, which strongly ''blueshift'' the spectra. On the other hand, spectral broadening becomes enhanced by harmonic generation. 
The coupling of TH with an infrared (IR) pump produces a ''two-colored'' filament from pump intensities above 10~TW/cm$^2$
\cite{Yang:pre:67:015401,Akozbek:prl:89:143901,Alexeev:ol:30:1503}. The amount of pump energy transferred into TH 
radiation depends on the linear wave vector mismatch
parameter $\Delta k = [3k(\omega)- k(3\omega)]^{-1}$ fixing the
coherence length $L_c = \pi/|\Delta k|$. The smaller the coherence
length, the weaker TH fields. Along meter-range distances, the
TH component can stabilize the pump wave with about $0.5
\%$ conversion efficiency \cite{Berge:pre:71:016602}. Experimental and numerical
data reported ring structures embarking most of the TH energy and
having a half-divergence angle of about 0.5~mrad
\cite{Theberge:oc:245:399}. This process contributes to create a continuous spectral band of UV-visible wavelengths \cite{Akozbek:apb:77:177,Berge:pre:71:016602,Mejean:apb:82:341}. 

Resembling spectral dynamics have also been reported from 1-mJ
infrared pulses propagating in argon at atmospheric pressure, after
subsequent compression by chirped mirrors \cite{Trushin:apb:80:399}.
Simulations of these experiments \cite{Akozbek:njp:8:177}, discarding TH emission, revealed that temporal gradients inherent to the steepening
operators are sufficient to amplify UV shifts and cover the TH
bandwidth down to 250 and 210~nm for initial pulse durations of 10 and 6~fs,
respectively. Very recently, numerical
simulations \cite{Kolesik:apb:85:531} refound this tendency for atmospheric propagation, i.e., TH generation, while it affects the pump dynamics to some extent over long ranges, does not
change significantly SC spectra, whose variations are mostly induced by 
the fundamental field in air.

Despite these last results, we are still missing a detailed
understanding of the key parameters which are supposed to drive SC
generation. A first important parameter is, of course, the laser
wavelength itself: How does the supercontinuum evolve when $\lambda_0$ is varied? This question was addressed in Ref.~\cite{Akozbek:apb:77:177} for various laser wavelengths, at which some spectral components were seen to merge. However, the model used a two-envelope approximation (for the pump and TH fields, separately). As emphasized in \cite{Kolesik:apb:85:531}, splitting into TH and SC pump within envelopes becomes problematic when their respective spectra overlap inside a
wide frequency interval where the basic validity condition $\Delta \omega_j/\omega_j \ll
1$ ($j = \omega, 3\omega$) may no longer be fulfilled. Actually, TH radiation produced through the nonlinear polarization needs to be described self-consistently from a single equation governing the total real
optical field. This model was missing in Refs.~\cite{Berge:pre:71:016602,Mejean:apb:82:341}, which
made the role of TH overestimated in the white light emission. 
Another important parameter is the length of the self-guiding range:
Successive cycles of focusing and
defocusing events promote the creation of shorter peaks in
the pulse temporal profile and lead to a maximal extension of the
spectrum.
A third potential player is the input pulse duration. In \cite{Akozbek:njp:8:177}, this was shown to affect the spectra in noble gases
for pulses containing a few optical cycles mainly. Clearing this aspect
requires several simulations using distinct pulse durations and
exploiting different propagation ranges. In connection, 
we demonstrate that spectral enlargements are directly linked to the level of maximum intensity:
Steepening operators as well as TH radiation broaden all the more the spectra as the intensity in the filament is high.

The paper is organized as follows: Sec.~\ref{Model_equations} presents the model
equations, namely, a unidirectional propagation equation for the total
electric field that generates higher-order harmonics (mostly TH)
through Kerr nonlinearities. Results from this equation will be
compared with those inferred from the ''standard'' nonlinear evolution
equation (NEE) for the pump wave. The major difference between these two models lies in the production of the TH field and its coupling with the pump
wave. Sec.~\ref{Long-range} is devoted to the long-range propagation of 127-fs
pulses in air described by the previous models. 
Emphasis is put on the influence of the central wavelength $\lambda_0$
(248, 800, 1550~nm). We discuss spectral modifications versus the
height of $I_{\rm max}$, the input duration, together 
with the temporal steepening dynamics and merging between TH
and pump spectral bands. Sec.~\ref{Short-range} revisits SC
for short-range (focused) propagations. It is shown that $I_{\rm max}$ becomes
closer to analytical evaluations when the beam develops few focusing/defocusing cycles. In this configuration, a lesser broadening may be achieved. Sec.~\ref{conclusion} finally summarizes the
generic features resulting from our analysis.

\section{Models for pulse propagation and underlying
physics}

\label{Model_equations}

Our unidirectional pulse propagation equation (UPPE) assumes scalar and
radially-symmetric approximations. It also supposes negligible
backscattering. These hypotheses hold as long as the beam keeps
transverse extensions larger than the central laser wavelength and as the
nonlinear responses (together with their longitudinal variations) 
are small compared with the linear refraction index. 
Straightforward manipulations of Maxwell equations allow us to
establish the equation for the spectral amplitude of the optical
electric field in the forward direction as \cite{Berge:review}
\begin{equation}
\label{UPPE_1}
\partial_z \widehat{E} = \frac{i}{2 k(\omega)} \nabla_{\perp}^2
\widehat{E} + i k(\omega) \widehat{E} + \frac{i \mu_0 \omega^2}{2
k(\omega)} \widehat{\cal F}_{\rm NL},
\end{equation}
where $\widehat{E}(r,z,\omega) = (2\pi)^{-1} \int E(r,z,t) \mbox{e}^{i
\omega t} dt$ is the Fourier transform of the forward electric field
component, $z$ is the propagation variable, $\nabla_{\perp}^2 = r^{-1}
\partial_r r \partial_r$ ($r \equiv \sqrt{x^2+y^2}$) is the
diffraction operator, $\mu_0 \epsilon_0 = 1/c^2$, $k(\omega) =
\sqrt{1 + \chi^{(1)}(\omega)} \omega/c$ is the wavenumber of the optical
field depending on the linear susceptibility tensor $\chi^{(1)}(\omega)$
defined at frequency $\omega$. In Eq.~(\ref{UPPE_1}), $\widehat{\cal
F}_{\rm NL} \equiv \widehat{P}_{\rm NL} + i\widehat{J}/\omega$ is the
Fourier transform of the nonlinearities that include the nonlinear
optical polarization $P_{\rm NL}$ and the current density $J$ created by
charged particles. Eq.~(\ref{UPPE_1}) restores the earlier UPPE
formulation proposed by Kolesik {\it et al.}~\cite{Kolesik:prl:89:283902} in the limit
$k_{\perp}^2/k^2(\omega) \ll 1$ ($k_{\perp}^2 = k_x^2 + k_y^2$).

For practical use, it is convenient to introduce the complex version
of the electric field
\begin{equation}
\label{complex}
E = \sqrt{c_1}(\mathcal{E} + \mathcal{E}^*),\,\,\,\mathcal{E} =
\frac{1}{\sqrt{c_1}} \int \Theta(\omega) \widehat{E} \mbox{e}^{- i
\omega t} d\omega,
\end{equation}
where $c_1 \equiv \omega_0 \mu_0/2 k_0$ employs the central wavenumber and
frequency of the pump wave ($k_0 \equiv n_0 \omega_0/c$) and $\Theta(x)$ denotes the Heaviside function. Because $\mathcal{E}$ satisfies
$\widehat{\mathcal{E}^*}(\omega) = \widehat{\mathcal{E}}(-\omega)^*$
($^*$ means complex conjugate), it is then sufficient to treat the
UPPE model~(\ref{UPPE_1}) in the frequency domain $\omega > 0$ only. 
The field intensity can be defined by $E^2$ averaged over an optical
period at least. Expressed in W/cm$^2$, it is simply given by the classical
relation $I = |\mathcal{E}|^2$.

Concerning the nonlinearities, we assume a linearly
polarized field. We consider a cubic susceptibility tensor
$\chi^{(3)}$ keeping a constant value around $\omega_0$, so that $P_{\rm NL}$ contains the instantaneous cubic polarization 
expressed as $P^{(3)}(\vec{r},t) = \epsilon_0 \chi_{\omega_0}^{(3)} E^3$. In addition, the phenomenon of
Raman scattering comes into play when the laser field interacts with
anisotropic molecules, in which vibrational and rotational states are
excited. Depending on the transition frequency $\Omega_{13}$ in
three-level molecular systems and
related dipole matrix element $\mu$ \cite{Penano:pre:68:056502}, the Raman response
takes the form
\begin{equation}
\label{PRaman}
P_{\rm Raman}
= \frac{2 \chi^{(1)} \mu^2}{\Omega_{31} \hbar^2} \int_{-\infty}^t
\mbox{e}^{-\frac{t-t'}{\tau_2}} \sin(\frac{t-t'}{\tau_1}) E^2(t') dt' \times E,
\end{equation}
where $\tau_1 = 1/\omega_R$ is the inverse of the fundamental
rotational frequency and $\tau_2$ is the dipole dephasing time.
Expressed in terms of the rescaled complex field $\mathcal{E}$ 
[Eq.~(\ref{complex})] and with appropriate normalizations \cite{Sprangle:pre:66:046418}, Eq.~(\ref{PRaman})
completes the cubic polarization as
\begin{subequations}
\label{Raman_123}
\begin{align}
\label{Raman_1}
\begin{split}
P_{\rm NL} & = 2 n_0 n_2 \epsilon_0 \sqrt{c_1} \int_{-\infty}^{+\infty}
{\bar R}(t-t') |\mathcal{E}(t')|^2 dt' \mathcal{E} \\
& \quad + 2 n_0 n_2 \epsilon_0 \sqrt{c_1} (1 - x_K)
\mathcal{E}^3/3+c.c.,
\end{split} \\
\label{Raman_3}
{\bar R}(t) & = (1 - x_K) \delta(t) + x_K \Theta (t) h(t), \\
\label{Raman_2}
h(t) & = \frac{\tau_1^2 + \tau_2^2}{\tau_1 \tau_2^2}
\mbox{e}^{-t/\tau_2} \sin(t/\tau_1),
\end{align}
\end{subequations}
where $n_2= 3
\chi_{\omega_0}^{(3)}/(4 n_0^2 c \epsilon_0)$ is the Kerr nonlinear index. Expression~(\ref{Raman_1})
possesses both retarded and instantaneous components
in the ratio $x_K$. The instantaneous part $\sim \delta(t)$ of Eq.~(\ref{Raman_3}) describes
the response from the bound electrons. The retarded part
$\sim h(t)$ accounts for the
Raman contribution, in which fast oscillations in $E^2$ give
negligible contributions, as $\tau_1 \sim \tau_2 \sim 70$~fs
are assumed to exceed the optical period $\sim \omega_0^{-1}$.

When free electrons are created, they induce a current density
$J = q_e \rho v_e$, which depends on the electron charge
$q_e$, the electron density $\rho$ and the electron velocity $v_e$. 
$J$ is computed from fluid equations involving external plasma sources
and the electron collision frequency $\nu_e$. At moderate
intensities ($< 10^{15}$~W/cm$^2$), the current density obeys
\begin{equation}
\label{plasma_3}
\partial_t J + \nu_e J = \frac{q_e^2 \rho}{m_e} E.
\end{equation}
Assuming electrons born at rest, the growth of the electron density is
only governed by external source terms, i.e.,
\begin{equation}
\label{rho1}
\partial_t \rho = W(I) (\rho_{\rm nt} - \rho) + \frac{\sigma}{U_i}
\rho I,
\end{equation}
that include photo-ionization processes with rate $W(I)$ and
collisional ionization with cross-section
\begin{equation}
\label{sigma}
\sigma(\omega) = \frac{q_e^2}{m_e \epsilon_0 n_0 c \nu_e (1 +
\omega^2/\nu_e^2)}.
\end{equation}
Here, $\rho_{\rm nt}$ and
$U_i$ are the density of neutral species and the ionization potential,
respectively. Electron recombination
in gases is efficient over long (ns) time scales, and
therefore we omit it. In Eq.~(\ref{rho1}), the rate for photo-ionization $W(I)$
follows from the Perelomov, Popov and Terent'ev
(PPT)'s theory \cite{Perelomov:spjetp:24:207} incorporating Ammosov, Delone and Krainov
(ADK) coefficients \cite{Ammosov:spjetp:64:1191} (see also Ref.~\cite{Nuter:josab:23:874}). Optical
field ionization theories stress two major limits bounded by the
Keldysh parameter,
\begin{equation}
\label{keldyshparam}
\gamma = \omega_0 \frac{\sqrt{2 m_e U_i}}{|q_e| E_p},
\end{equation}
namely, the limit for Multi-Photon Ionization (MPI, $\gamma \gg 1$)
concerned with rather low intensities and the tunnel limit ($\gamma
\ll 1$) concerned with high intensities, from which the Coulomb
barrier becomes low enough to let the electron tunnel out. Here,
$E_p$ denotes the peak optical amplitude. 
For laser intensities $I = |\mathcal{E}|^2 < 10^{13}-10^{14}$~W/cm$^2$, MPI
characterized by the limit
\begin{equation}
W(I) \rightarrow W_{\rm
MPI} = \sigma_K I^K
\label{mpi}
\end{equation}
dominates, where $K=\textrm{mod}(U_i/\hbar\omega_0)+1$ is the number
of photons necessary to liberate one electron. The level of clamped 
intensity, $I_{\rm max}$, depends on the selected ionization
rate. 

Energy lost by the pulse through single ionization processes is 
determined by a local version of the Poynting theorem, yielding the 
loss current $J_{loss}$, such that $J_{loss} \cdot E = U_i 
W(I)(\rho_{\rm nt} - \rho)$. As a result, our UPPE model reads in Fourier space as
\begin{equation}
\label{finalUPPE1}
\begin{split}
\frac{\partial}{\partial z} \widehat{\cal E} & = \left[ \frac{i}{2
k(\omega)} \nabla_{\perp}^2 + i k(\omega) \right] \widehat{\cal E} + \frac{i
\mu_0 \omega^2}{2 k(\omega) \sqrt{c_1}} \Theta(\omega) \widehat{P}_{\rm NL}
\\
 & - \frac{i k_0^2 \Theta(\omega)}{2
\epsilon(\omega_0) k(\omega)(1+\frac{\nu_e^2}{\omega^2})}
\left(\widehat{\frac{\rho
\mathcal{E}}{\rho_c}}\right) - \frac{\Theta(\omega)}{2}
\sqrt{\frac{\epsilon(\omega_0)}{\epsilon(\omega)}} {\cal L}(\omega),
\end{split}
\end{equation}
where
\begin{equation}
\label{TFlosses}
{\cal L}(\omega) = \frac{U_i}{2\pi} \int \mathcal{E} \left[
\frac{W(I)}{I}(\rho_{\rm nt} - \rho) + \frac{\sigma(\omega)}{U_i} \rho
\right] \mbox{e}^{i \omega t} dt.
\end{equation}
In Eq.~(\ref{finalUPPE1}), $P_{\rm NL}$ and the expression containing the electron density $\rho(\vec{r},t)$ [Eq.~(\ref{rho1})] must be
transformed to Fourier space, from which we retain only positive frequencies for the symmetry reasons given above.

Alternatively, when a central frequency $\omega_0$ is imposed,
Eq.~(\ref{UPPE_1}) restitutes the Nonlinear Envelope Equation (NEE),
earlier derived by Brabec and Krausz \cite{Brabec:prl:78:3282}. We can make use of the Taylor expansion
\begin{equation}
\label{taylor}
k(\omega) = k_0 + k' \bar{\omega} + \widehat{\cal
D},\,\,\, \widehat{\cal D} \equiv \sum_{n \geq 2}^{+\infty}
\frac{k^{(n)}}{n!} \bar{\omega}^n,
\end{equation}
where $\bar{\omega} = \omega - \omega_0$, $k' = \partial k/\partial
\omega|_{\omega = \omega_0}$, $k^{(n)} = \partial^n k/\partial
\omega^n|_{\omega = \omega_0}$, and take the inverse Fourier transform of
Eq.~(\ref{UPPE_1}) in which terms with $k(\omega)$ in their
denominator are expanded up to first order in $\bar{\omega}$ only.
After introducing the complex-field representation $\mathcal{E} = U
\mbox{e}^{i k_0 z - i \omega_0 t}$, the new time variable $t \rightarrow t -
z/v_g$ can be utilized to replace the pulse into the frame moving with
the group velocity
$v_g= k'^{-1}$. Furthermore assuming $\nu_e^2/\omega_0^2
\ll 1$, $\sqrt{\epsilon(\omega_0)/\epsilon(\omega)}
\approx 1$ and ignoring the TH component, the nonlinear envelope
equation for the forward pump envelope $U$ expands as follows:
\begin{equation}
\label{1}
\begin{split}
\frac{\partial}{\partial z} U & = \frac{i}{2k_0}
{T}^{-1}\nabla_{\perp}^2 U + i \mathcal{D} U + i \frac{\omega_0}{c}
n_2 T \times \\
 & \quad \left[ \left(1-x_K \right)|U|^2 + x_K \int_{-\infty}^t
 h(t-t') \left|U(t^{\prime})\right|^2
dt^{\prime} \right] U \\
 & \quad - i \frac{k_0}{2n_0^2 \rho_c} {T}^{-1} \rho U -
 \frac{\sigma}{2} \rho U - (\rho_{\rm nt} - \rho)\frac{U_i W(I)}{2|U|^2} U,
\end{split}
\end{equation}
where $\sigma = \sigma(\omega_0)$, ${\cal D} \equiv \sum_{n \geq 2}^{+\infty} (k^{(n)}/n!) (i 
\partial_{t})^n$ and $T = (1 + \frac{i}{\omega_0}
\partial_{t})$. The first term of the operator $\mathcal{D}$
corresponds to group-velocity dispersion with coefficient $k''= \partial^2
k/\partial \omega^2|_{\omega = \omega_0}$. Equation~(\ref{1})
describes wave diffraction, Kerr focusing response,
plasma generation, chromatic dispersion with self-consistent deviations 
from the classical slowly-varying envelope approximation through the space-time focusing and self-steepening operators [$({T}^{-1}
\nabla_{\perp}^2 \mathcal{E})$ and $(T |\mathcal{E}|^2
\mathcal{E})$, respectively]. This model will be integrated
numerically by using initially Gaussian pulses,
\begin{equation}
\label{incond}
U(x,y,z=0,t) = \sqrt{\frac{2P_{\rm in}}{\pi w_0^2}}
\mbox{e}^{-\frac{r^{2}}{w_0^{2}} - i k_0
\frac{r^2}{2f} - \frac{t^2}{t_p^2}},
\end{equation}
which involves the input power $P_{\rm in}$, the beam waist $w_0$ and
$1/e^2$ pulse half-width $t_p$. Input pulses can be focused
through a lens of focal length $f$ and they linearly diffract over the distance
\begin{equation}
\label{zf}
z_f = (f^2/z_0)/(1 + f^2/z_0^2),
\end{equation}
where $z_0 = \pi n_0 w_0^2/\lambda_0$ is the Rayleigh
range of the collimated beam ($f = +\infty$).

\begin{figure}
\includegraphics[width=\columnwidth]{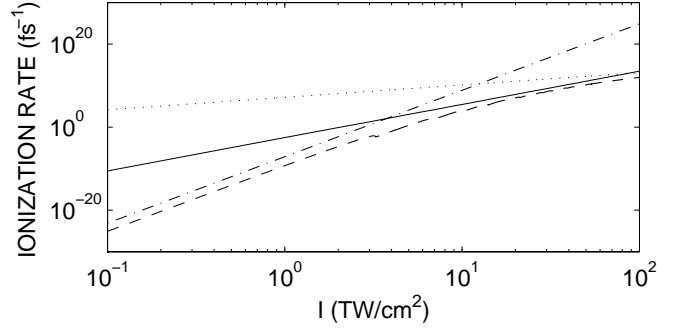}
\caption{\label{fig0} Ionization rates for the different wavelengths used throughout this paper:
$\lambda_0=800$~nm (solid line), $\lambda_0=248$~nm (dotted line), and $\lambda_0=1550$~nm (dashed line).
The dashed-dotted line shows the overestimated ionization rate chosen in Sec.~\ref{imax} for $\lambda_0=1550$~nm.}
\end{figure}

In the coming analysis, we shall employ the nonlinear refractive indices $n_2=8\times10^{-19}$, 
$4\times10^{-19}$, and $1\times10^{-19}$~cm$^2$/W for the wavelengths $\lambda_0 = 248$, 800 and 1550~nm,
respectively. At 800~nm, we consider a fitted MPI formulation for the ionization rate, $W(I) \rightarrow \sigma_{(K)} I^K$, where $K = 8$ and $\sigma_{(8)}=2.88\times 10^{-99}$~s$^{-1}$cm$^{16}$/W$^8$.
This approximation is known to reproduce experimental data at $\lambda_0 = 800$ nm rather faithfully \cite{Nuter:josab:23:874,Skupin:pre:70:046602}. For the two other wavelengths, we lack well established formulations, so we employ PPT ionization rates. We consider $O_2$ molecules having the lowest gap potential ($U_i = 12.1$ eV) as the main specy undergoing
ionization with an effective residual charge $Z_{\rm eff}=0.53$ \cite{Talebpour:oc:163:29}. All ionization rates used in the present paper are illustrated in Fig.~\ref{fig0}. They yield saturation intensity below the
threshold of 100 TW/cm$^2$ currently claimed in the literature
\cite{Berge:review,Akozbek:apb:77:177}. The dispersion relation for air has been parametrized as in Ref.~\cite{Peck:josa:62:958}.

Since our outlook is to understand spectral variations versus the propagation dynamics at different wavelengths, we find it instructive to fix the same ratio of input power over critical, e. g., $P_{\rm in} = 4 \times P_{\rm cr}$, at all wavelengths. To locate the Kerr-driven filamentation onset upon comparable $z$ scales, we also adjust the ratio of the nonlinear focus $z_c$ and Rayleigh range $z_0$ between 2 and 4, by adapting suitably the beam waist $w_0$ between 1 and 4~mm.

\subsection{Self-phase modulation and SC generation}

In air, dispersion is weak with $k'' \lesssim 1$ fs$^2$/cm in the UV as well as in the mid-IR domains. Self-channeling then mainly
relies on the dynamical balance between Kerr self-focusing and plasma
defocusing, so that estimates for peak intensities ($I_{\rm max}$), electron densities ($\rho_{\rm max}$) and filament radius ($L_{\rm min}$) can be deduced 
from equating diffraction, Kerr and ionization responses in Eq. (\ref{1}). This yields the simple relations
\begin{subequations}
\label{estimate13}
\begin{gather}
\label{estimate1}
I_{\rm max} \approx \frac{\rho_{\rm max}}{2 \rho_c n_0 {\bar 
n}_2}, \quad
\rho_{\rm max} \approx t_p \rho_{\rm nt} W(I_{\rm max}), \\
\label{estimate3}
L_{\rm min} \approx \pi (2 k_0^2 {\bar n}_2 I_{\rm max}/n_0)^{-1/2},
\end{gather}
\end{subequations}
where
\begin{equation}
\label{n2bar}
{\bar n}_2 = n_2 (1 - x_K) + n_2 x_K \mbox{max}_t \int_{-\infty}^t h(t-t') \textrm{e}^{-2\frac{t'^2}{t_p^2}} dt',
\end{equation}
represents the maximal effective Kerr index over the initial pulse profile.
For practical use, $W(I_{\rm max})$ can be simplified to $\sigma_K I_{\rm max}^K$ in MPI-like formulation.

The magnitude of $I_{\rm max}$ directly impacts spectral broadening, which 
is basically driven by self-phase modulation (SPM). Because the frequency spectrum is expanded by the nonlinearity, SPM leads to SC, as the wave 
intensity strongly increases through the self-focusing
process. Noting by $\varphi(\vec{r},t)$ the phase of the field envelope, frequency variations are dictated in the limit $T, T^{-1} \rightarrow 1$ by
\begin{equation}
\label{SPMChin}
\Delta \omega = - \partial_t \varphi \sim - k_0 \Delta z \partial_t
({\bar n_2} I - \rho/2 n_0 \rho_c),
\end{equation}
which varies with the superimposed actions of the Kerr and plasma
responses. Near the focus point $z_c$, only the front edge of the pulse survives
from this interplay and a redshift is enhanced by plasma
generation. At later distances, second focusing/defocusing sequences
attenuate this first tendency. In contrast, when accounting for temporal steepening ($T, T^{-1} \neq 1$), shock
edges in the back of the pulse are created and a ''blue shoulder'' appears in the spectrum, to the detriment of the early redshift \cite{Berge:review,Akozbek:oc:191:353,Rothenberg:ol:17:584,Yang:ol:9:510}.

In addition, the cubic polarization generates third-order
harmonics, modeled by the last term of Eq.~(\ref{Raman_1}). 
In self-focusing regimes, the third-harmonic intensity usually
contributes by a little percentage to the overall beam fluence
\cite{Akozbek:prl:89:143901}. Despite its smallness, this
component may act as a saturable nonlinearity for the carrier wave. 
It lowers the peak intensity of the pump and contributes to enhance the blue side of the spectrum after the TH and pump bandwidths increase and overlap \cite{Akozbek:apb:77:177,Berge:pre:71:016602}.

\section{Long-range Propagation}

\label{Long-range}

We numerically analyze supercontinuum generation for the three
laser wavelengths of 248 nm, 800 nm and 1550 nm. Results of
the nonlinear Schr{\"o}dinger-like equation~(\ref{1}) for the pump envelope involving or not space-time focusing and
self-steepening operators are compared with those of the unidirectional propagation 
equation~(\ref{finalUPPE1})
avoiding any Taylor expansion in the dispersion relation.

\begin{figure}
\includegraphics[width=\columnwidth]{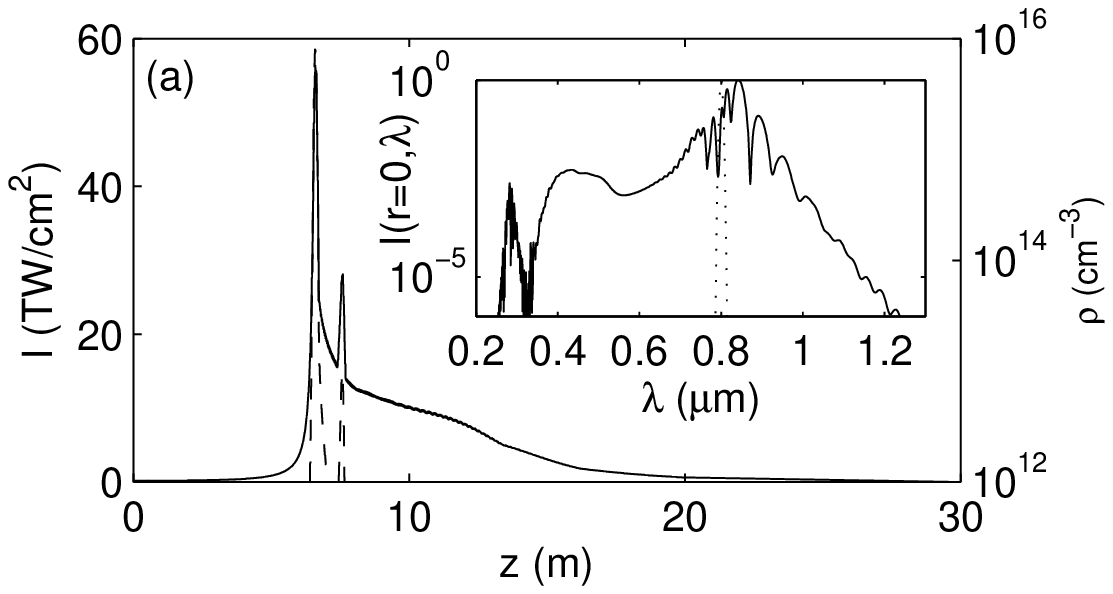}
\includegraphics[width=\columnwidth]{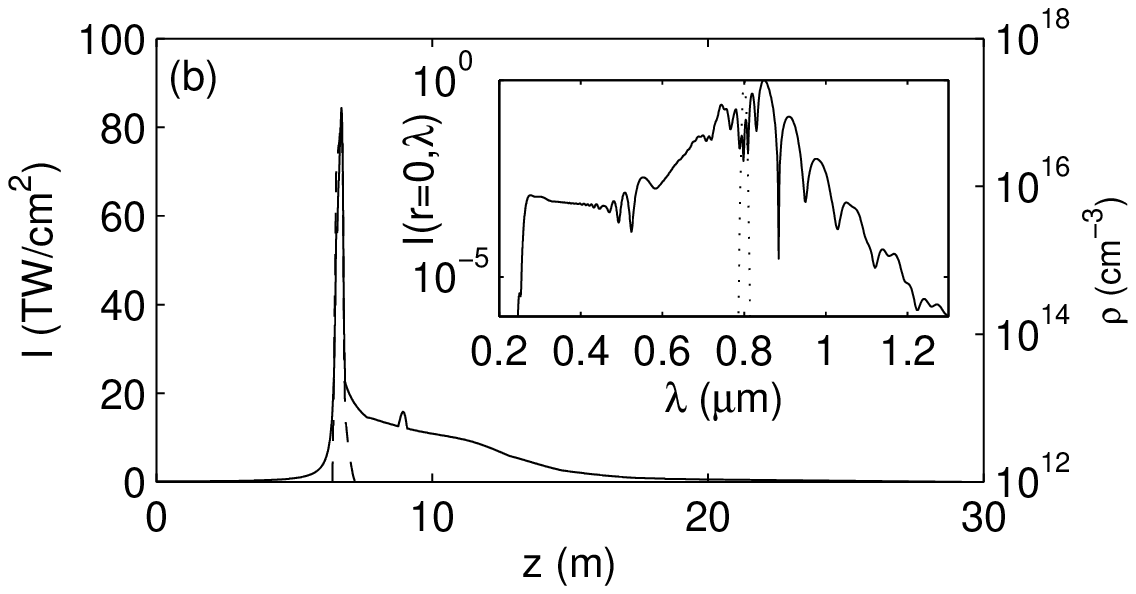}
\includegraphics[width=\columnwidth]{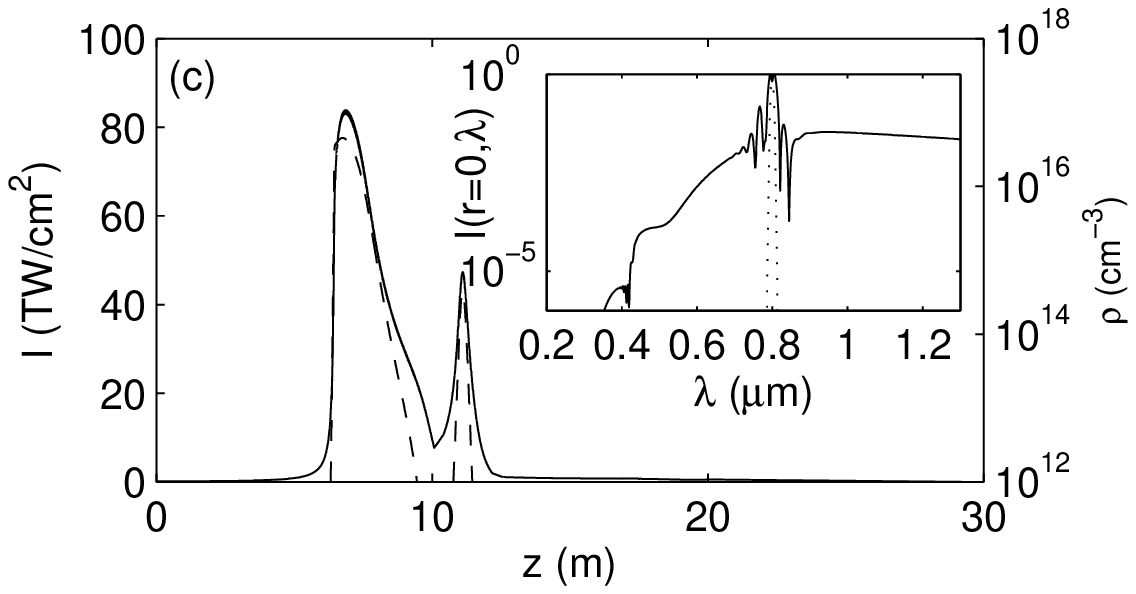}
\caption{\label{fig1} Peak intensities (solid curves, left-hand side scales) and peak electron densities (dashed curves, right-hand side scales) of 127-fs pulses with ratio of 
input power over critical equal to 4 and waist $w_0 = 2 {\rm~mm}$ at $\lambda_0=800$~nm.
The insets show on-axis spectra: dotted curves represent the initial spectra;
the solid curves the spectra at the propagation distance $z_{\rm max}=8$~m
where maximal broadening is observed: (a) UPPE model Eq.~(\ref{finalUPPE1}); (b) NEE Eq.~(\ref{1}) applied to the pump wave; (c) NEE Eq.~(\ref{1}) modified with setting $T=T^{-1} = 1$.}
\end{figure}

Special attention is here given to the long-range propagation, which 
rather favors several cycles of focusing-defocusing events. Before 
proceeding with the above parameters specifically, we perform three different series of simulations showing SC at 800 nm, whose results are summarized in Fig.~\ref{fig1}. The first one concerns direct integrations of 
Eq.~(\ref{finalUPPE1}); the second refers to the same pulse 
described by Eq.~(\ref{1}), which eludes TH production; the third 
approach relies on Eq.~(\ref{1}), in which temporal steepening is 
omitted, i.e., $T = T^{-1} = 1$. 

The insets in Fig.~\ref{fig1} show the on-axis spectra when SC is maximal. Following the UPPE description, TH, which emerges from $z_c \sim 6$~m, develops a limited redshift, whereas SC of the fundamental is widely extending towards the blue/UV wavelengths [Fig.~\ref{fig1}(a)]. Note that, although a broad plateau occurs in this domain, the TH bandwidth still appears
separated from the pump spectrum. Following the 
NEE description, there is no TH generation.
However, SC is so amplified in the blue region by 
temporal steepening effects, that it overlaps the TH zone and 
simply hides it [Fig.~\ref{fig1}(b)]. Finally, when neglecting temporal steepening, the pump instead develops a wide redshift (overestimated by plasma coupling) and a much narrower blueshift [Fig.~\ref{fig1}(c)].

A first observation can be drawn from Fig.~\ref{fig1}: Since TH is 
responsible for lowering the saturation intensity of the pump \cite{Berge:pre:71:016602}, $I_{\rm max}$ reached in the UPPE model is lower and the frequency variations $\Delta \omega \sim I_{\rm max} \Delta 
z/\Delta t$ are diminished compared with NEE spectra for the pump 
wave alone. Apart from this difference, no significant other change 
was detected between both these models, so that NEE seems to be nothing 
else but the UPPE description subtracted by the self-generated harmonics~\cite{Berge:review}. Importantly, 
omitting temporal derivatives of the operators $T, T^{-1}$ 
imply more serious discrepancies, as can be seen from Fig.~\ref{fig1}.

\subsection{Influence of $\lambda_0$}

\label{lambda}

\begin{figure}
\includegraphics[width=\columnwidth]{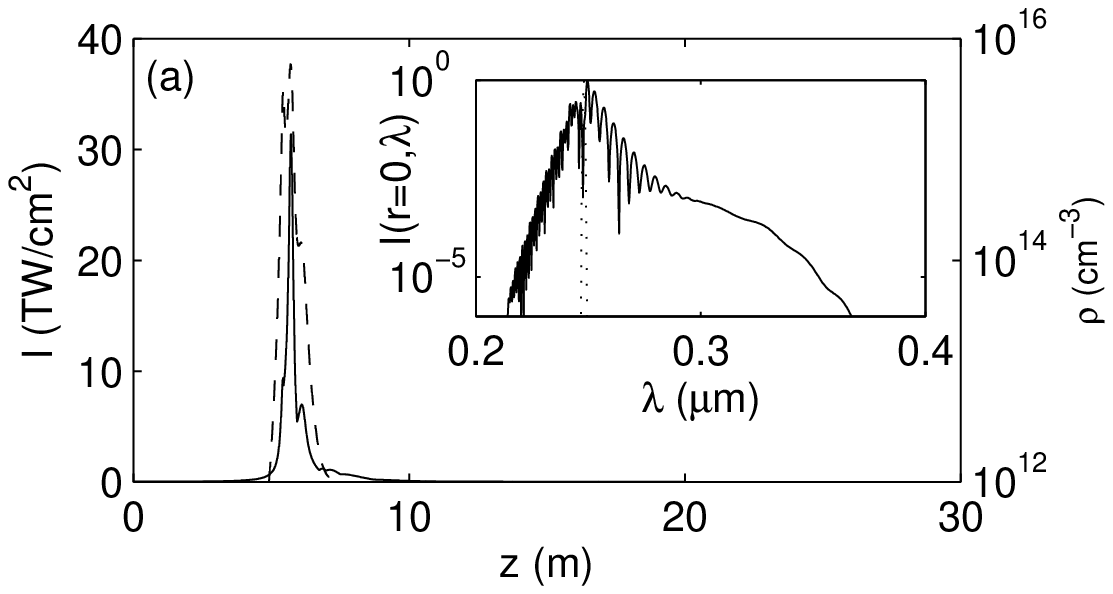}
\includegraphics[width=\columnwidth]{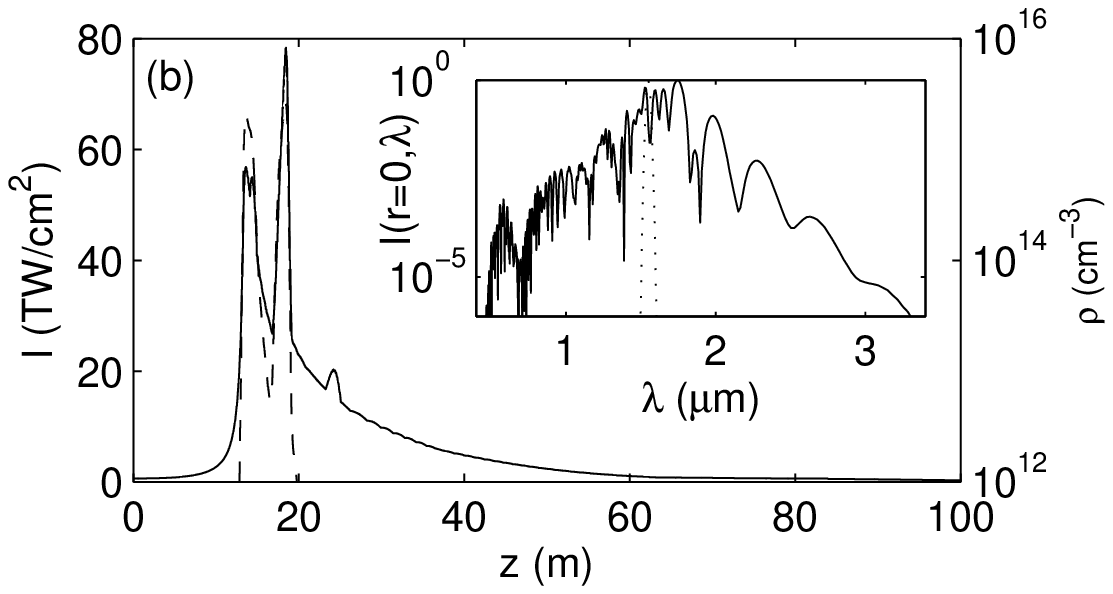}
\caption{\label{fig2} Peak intensities (solid curves, left-hand side scales) and peak electron densities (dashed curves, right-hand side scales) of 127-fs pulses with ratio of 
input power over critical equal to 4 at different
wavelengths $\lambda_0$. The insets show on-axis spectra: dotted curves represent the initial spectra;
the solid curves the spectra at the propagation distance $z_{\rm max}$
where maximal broadening is observed: (a) $\lambda_0=248$~nm, $w_0=1$~mm,
$z_{\rm max}=6$~m; (b) $\lambda_0=1550$~nm, $w_0=4$~mm, $z_{\rm max}=35$~m.}
\end{figure}

We now examine Gaussian pulses at different wavelengths (248, 800 and 1550~nm) with $t_p = 127$ fs and $P_{\rm in}/P_{\rm cr} = 4$  as 
initial conditions for the UPPE model in parallel geometry $(f =+\infty)$.
Figures~\ref{fig1}(a) and \ref{fig2} show snapshots of spectra at maximal extent, together with associated peak intensities and electron densities. We here specify that no TH generation was included for $\lambda_0 = 248$ nm, because no reliable data of the dispersion relation was available for this wavelength. We believe, instead, that spectral components below 90 nm should be rapidly absorbed by the medium.

\begin{figure}
\includegraphics[width=\columnwidth]{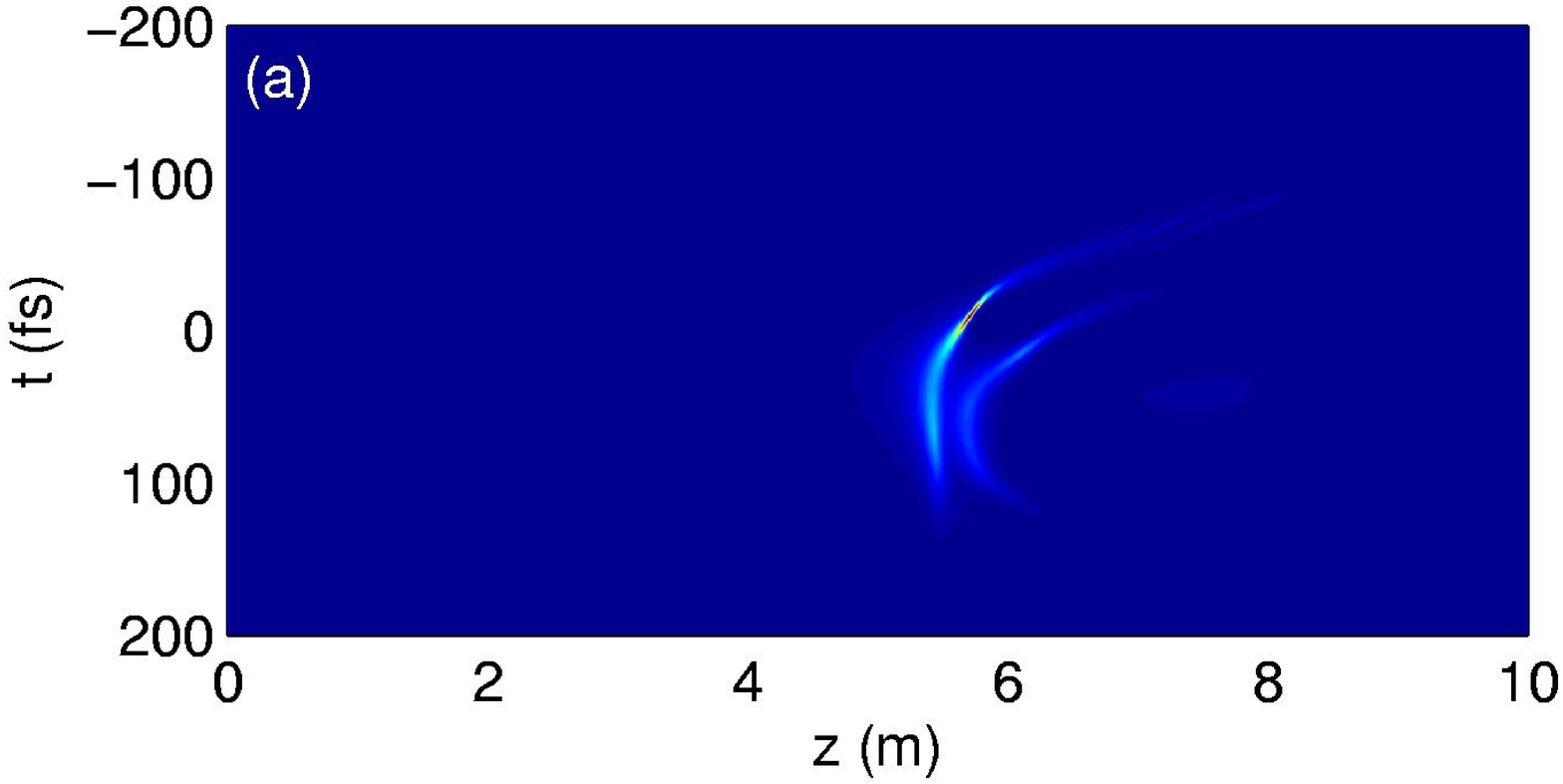}
\includegraphics[width=\columnwidth]{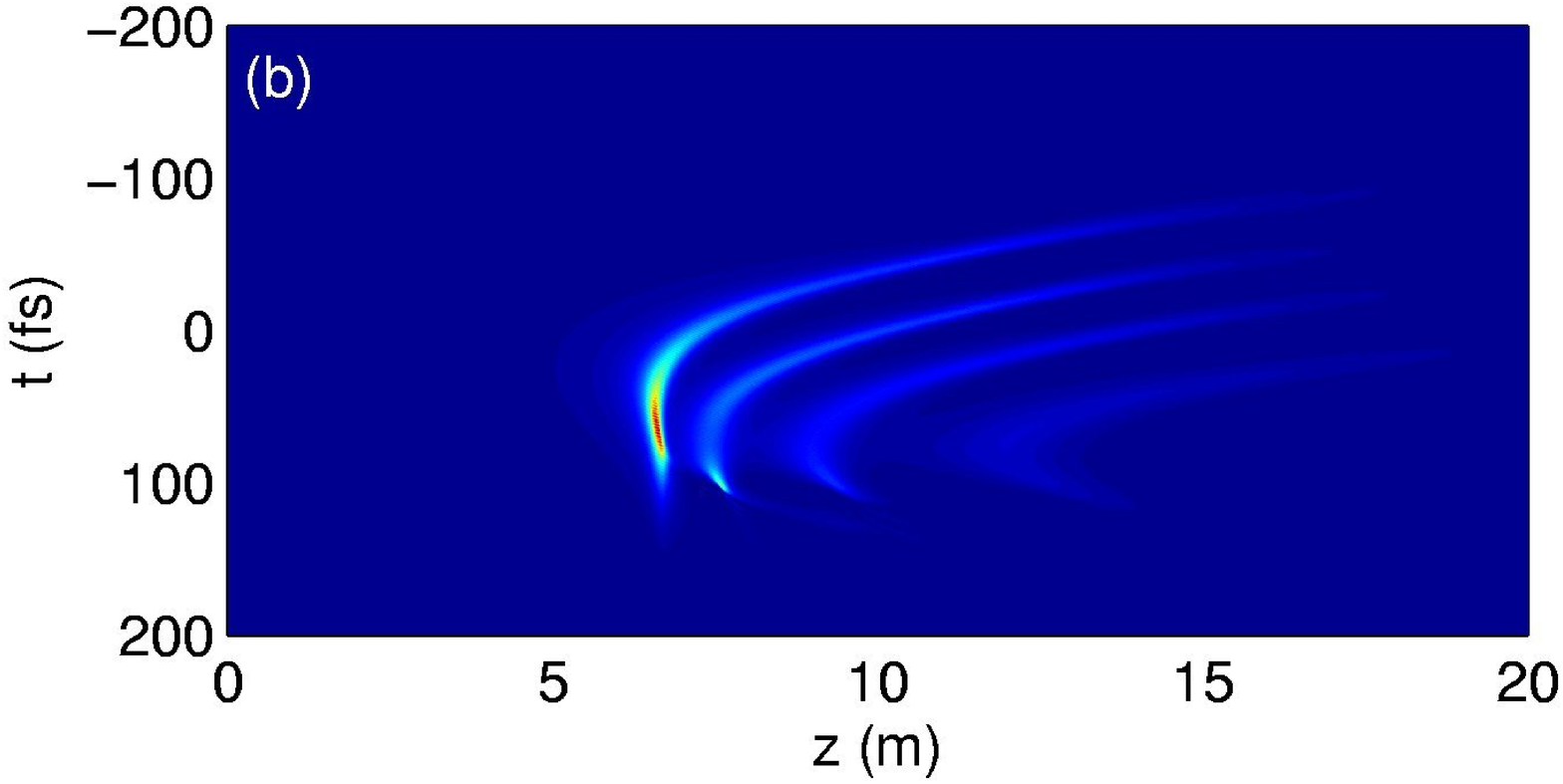}
\includegraphics[width=\columnwidth]{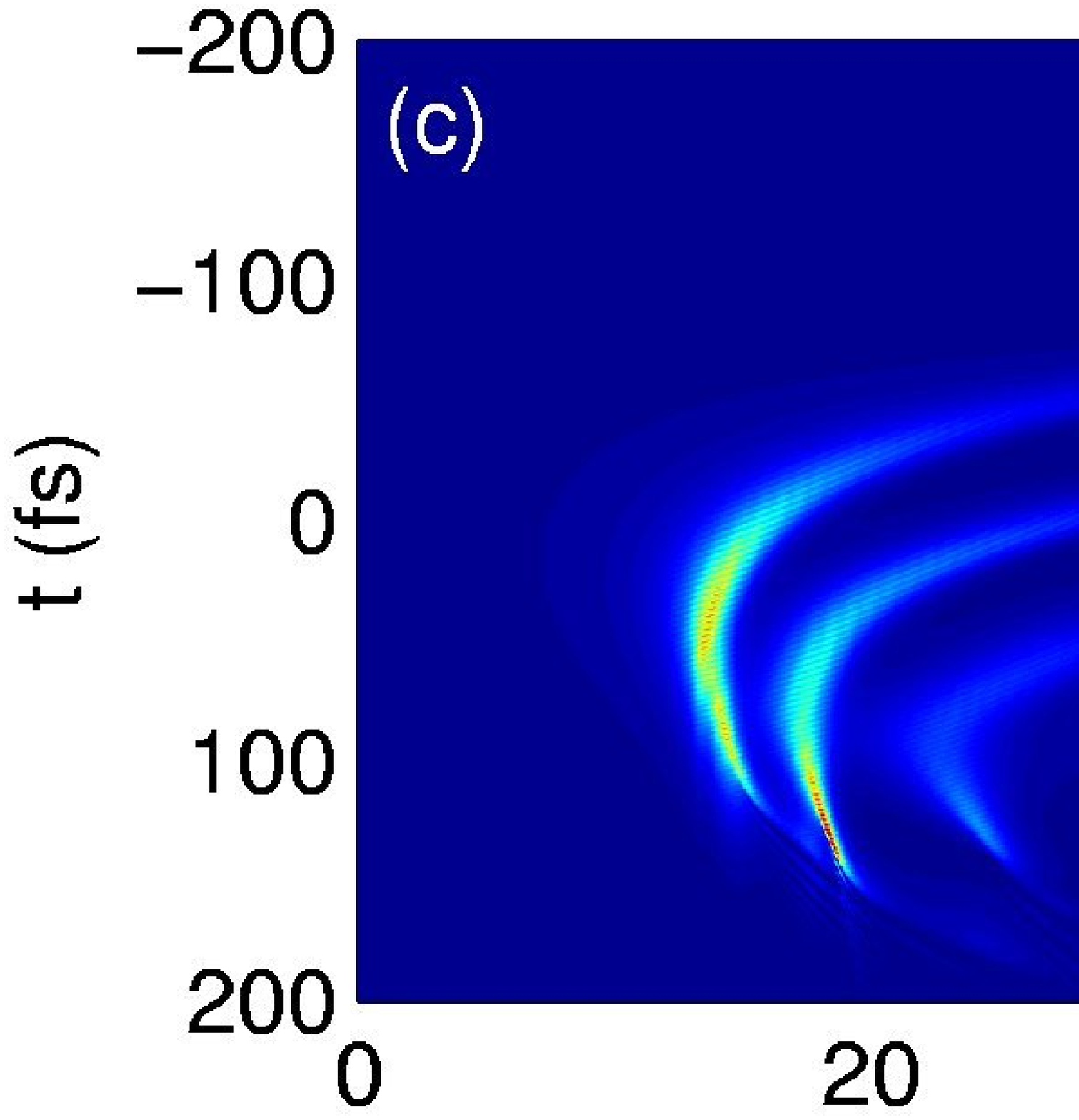}
\caption{\label{fig2bis} Temporal evolutions of 
the pulses shown in Figs.~\ref{fig1}(a) and \ref{fig2} in the $(t,z)$ plane: (a) $\lambda_0=248$~nm; (b) $\lambda_0=800$~nm; (c) $\lambda_0=1550$~nm.}
\end{figure}

By comparing Figs.~\ref{fig1}(a) and \ref{fig2}, it is seen right away that 
supercontinuum generation increases with the wavelength. To quantify this observation we introduce $\Delta\lambda_{\rm SC}$ as the total extension of the on-axis spectra over wavelengths at $10^{-5}$ times the maximal spectral intensity. Then a measurement for the effective broadening is the ratio $\Delta\lambda_{\rm SC}/\lambda_0$, which we find close to $\sim0.5$ at 248~nm, $\sim1$ at 800~nm and $\sim1.5$ at 1550~nm. A look at the propagation dynamics reveals that the self-guiding range is noticeably augmented at longer wavelengths. This can be explained by the transverse size of the filament. Equation~(\ref{estimate3}) gives an estimate for the beam waist in filamentation regime. If we assume comparable $I_{\rm max}$ for all wavelengths, we deduce that the filament diameter at 1550~nm is about one order of magnitude larger than that at 248~nm, which is compatible with our numerical data. Hence, the larger the wavelength, the slower the filament is expected to diffract.
Moreover, by virtue of the formula for the critical power $P_{\rm cr} \propto \lambda_0^2/n_2$ and since a filament conveys a few $P_{\rm cr}$ \cite{Berge:review}, it is obvious that IR filaments contain much more energy than their UV counterparts. Thus, nonlinear losses along the filamentation range are less dramatic in the IR domain. Due to the longer propagation range, more 
focusing/defocusing events participate in enlarging the spectra at longer wavelengths. Figure 4 details the evolution of the filaments in the plane $(t,z)$. It is seen that the time window in which the pulses disperse occupies the length of the input pulse duration. Although shorter temporal peaks arise through self-focusing/defocusing events, multi-peaked profiles mostly develop patterns having a whole extent close to $t_p$.

\begin{figure}
\includegraphics[width=\columnwidth]{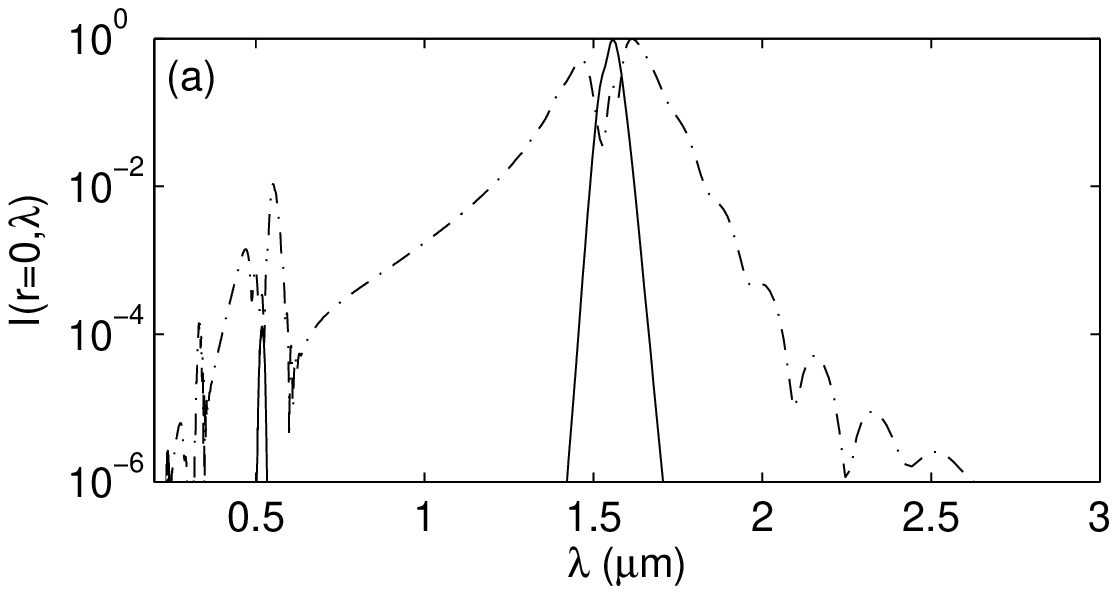}
\includegraphics[width=\columnwidth]{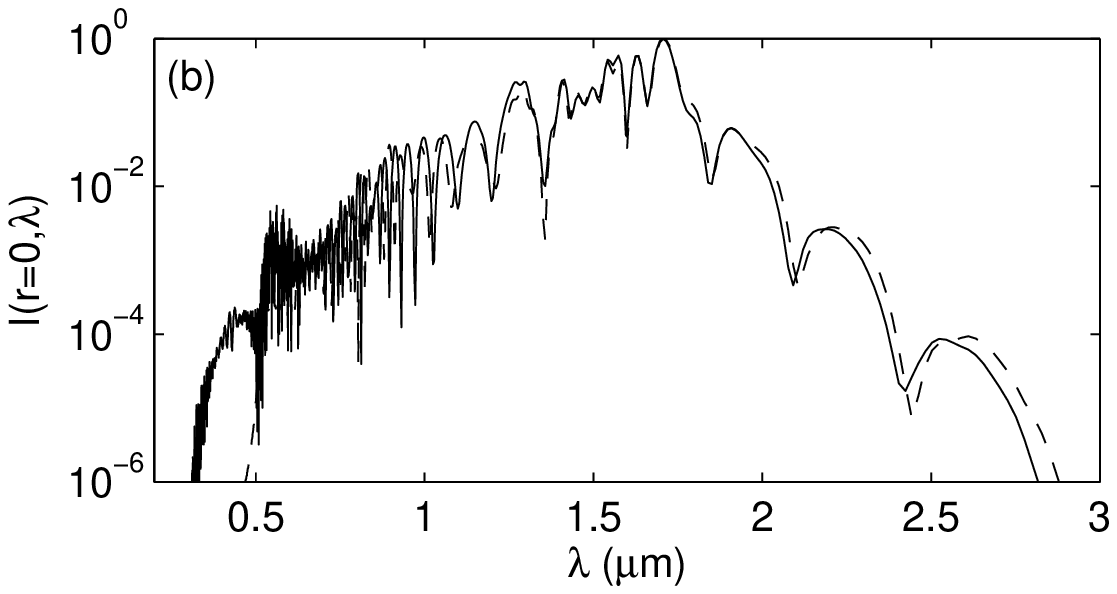}
\caption{\label{fig3} On-axis spectra for the pulse used in Fig.~\ref{fig2}(b) 
($\lambda_0 = 1550$~nm) for the propagation distances 
(a) $z = 10$~m (solid line) and  $z = 15$~m (dash-dotted line) and (b) $z = 25$~m (solid line). The dashed line in (b) refers to a spectrum 
at $z = 25$~m computed from the NEE model.}
\end{figure}

Before going on, we find it worth investigating SC at~1550 nm more thoroughly,
in relationship with third-harmonic generation. 
Figure~\ref{fig3} plots three of the SC development stages, first when 
TH and pump components are clearly separated ($z=10$~m), second when they start to merge ($z=15$~m).
At the last propagation distance ($z=25$~m), 
we can observe that, unlike Fig.~\ref{fig1}, TH and pump spectra overlap and make the TH bandwidth not 
distinguishable. For comparison, results from the NEE model for the 
pump wave alone have also been plotted.

\subsection{Influence of $I_{\rm max}$}

\label{imax}

\begin{figure}
\includegraphics[width=\columnwidth]{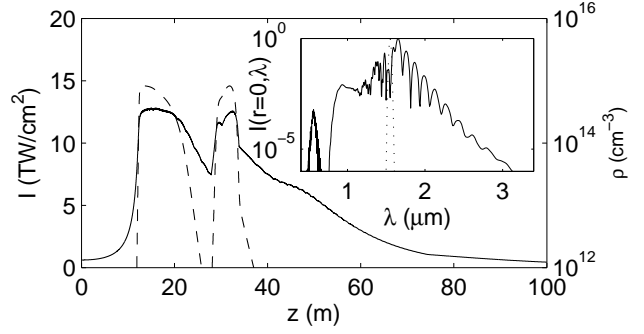}
\caption{\label{fig4} Peak intensity (solid curve, left-hand side scale) and peak electron density (dashed curve, right-hand side scale) for the same pulse as in Fig.~\ref{fig2}(b) ($\lambda_0 = 1550$~nm) computed from the UPPE model with an overestimated ionization rate (see Fig.~\ref{fig0}). The inset shows maximal spectral broadening attained at $z_{\rm max}=40$~m.}
\end{figure}

The impact of the saturation intensity onto SC is investigated by 
simply changing the ionization model: By decreasing the photo-ionization rate artificially it is possible to increase $I_{\rm max}$ and the maximal plasma level $\rho_{\rm max}$ accordingly. Reversely, increasing $W(I)$ reduces these two quantities, which can have a direct influence on the spectral broadening, as inferred from Eq.~(\ref{SPMChin}). To study this point,
we concentrate on the wavelength of 1550 nm only, because it yields the 
broadest spectra explored so far. Figure~\ref{fig4} shows the maximal intensity and peak electron density at this wavelength, when the ionization rate 
is artificially increased (see Fig.~\ref{fig0}). All other parameters are unchanged, compared with the simulation shown in Fig.~\ref{fig2}(b).
With the original ionization rate, $I_{\rm max}$ reaches the value of 80~TW/cm$^2$; with the artificial one, $I_{\rm max}$ stays below 13~TW/cm$^2$. The inset in Fig.~\ref{fig4} details the corresponding spectrum at $z_{\rm max}=40$~m. With low $I_{\rm max}$, the TH component is reduced to some extent as the pump intensity barely exceeds the TH conversion threshold \cite{Yang:pre:67:015401}. Meanwhile, SC of the pump driven by the $T, T^{-1}$ operators in the blue side decreases, i.e., a lower $I_{\rm max}$ for analogous pulse compression implies less sharp temporal gradients 
and smoother optical shocks, which weakens blueshifted frequency variations. These features are visible in Fig.~\ref{fig4}, where TH and pump broadbands remain separated. In Figs.~\ref{fig2}(b) and \ref{fig3}(b), in contrast, SC extends beyond the TH wavelength and increases more the 
pulse spectrum. Similar features were observed at the two other wavelengths, when the ionization rate was changed.

\subsection{Influence of $t_p$}

We investigate the influence of the initial pulse duration on the propagation dynamics and SC generation. Since we consider transform-limited pulses, the value of $t_p$ is directly linked to the initial spectral width. Moreover, $I_{\rm max}$ comes into play in SC generation and is expected to scale as $\sim (1/t_p)^{1/(K-1)}$ [see Eq.~(\ref{estimate1})]. Thus, the initial pulse duration should play a significant role in spectral broadening. To check this assessment, we performed several simulations using the UPPE model, by 
varying $t_p$ from 20~fs up to 500~fs. Because group-velocity dispersion becomes very efficient at short pulse durations and may even stop the Kerr self-focusing at powers too close to critical \cite{Skupin:pd:220:14,Skupin:pra:74:043813}, we increased the input power up to 20~$P_{\rm cr}$ for $t_p = 20$~fs. With this, we ensure to trigger a filamentation regime even for this short input duration.

\begin{figure}
\includegraphics[width=\columnwidth]{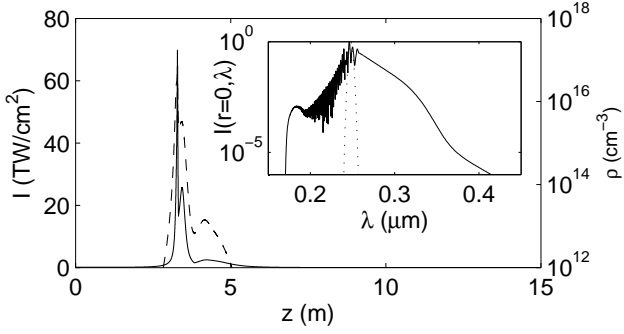}
\caption{\label{fig6} Peak intensity (solid curve, left-hand side scale) and peak electron density (dashed curve, right-hand side scale) of a 20-fs pulse with ratio of input power over critical equal to 20, $\lambda_0=248$~nm, $w_0=1$~mm. The inset shows on-axis spectra: the dotted curve represents the initial spectrum; the solid curve the spectrum at the distance $z_{\rm max}=3.5$~m where maximal broadening is observed.}
\end{figure}

To illustrate the dependency of $I_{\rm max}$ upon $t_p$, short wavelengths are preferable because the number of photons for ionization is small. Figure~\ref{fig6} shows maximum intensity, peak electron density and maximal spectral extent of a 20-fs pulse at 248~nm. At this wavelength, $K=3$ and, following Eq.~(\ref{estimate1}), $I_{\rm max}$ and $\rho_{\rm max}$ should increase
by a factor $\sim2.5$ compared to the 127-pulse shown in Fig.~\ref{fig2}(a). Indeed, both quantities are increased by a factor of $\sim 2$ in the simulation. Spectral broadening is augmented from 0.5 to 0.8 in terms of $\Delta \lambda_{\rm SC}/\lambda_0$, especially to the blue side. As explained in Sec.~\ref{imax}, this results from the action of the steepening operators. The overall propagation dynamics, characterized by the filamentation length and number of focusing/defocusing cycles are, however, comparable for both the 20-fs and the 127-fs pulses.

\begin{figure}
\includegraphics[width=\columnwidth]{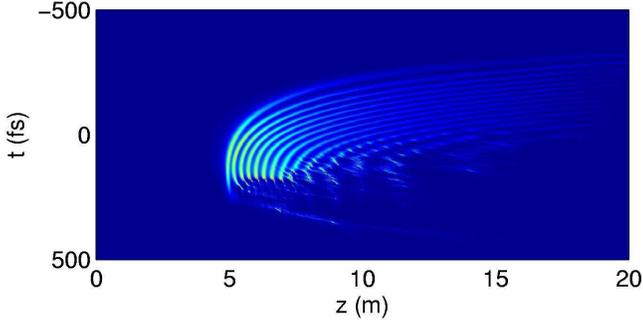}
\caption{\label{fig5bis2} Temporal evolution of a 500-fs pulse at 800~nm with 4~$P_{\rm cr}$ and waist $w_0 = 2 {\rm~mm}$ in the $(t,z)$ plane.}
\end{figure}

On the other hand, if we increase the initial duration $t_p$ towards the ps time scale, the propagation dynamics changes drastically. As an example, Figure~\ref{fig5bis2} shows the temporal evolution of a 500-fs pulse at 800~nm. Compared with Fig.~\ref{fig2bis}(b) employing $t_p = 127$ fs, the obvious difference is the huge number of focusing/defocusing cycles. The action of the generated plasma breaks the pulse profile into a larger number of shorter peaks. With a longer pulse duration, more ''time slices'' are available for feeding successive focusing events. The filamentation range is increased and $I_{\max}$ is maintained over several meters. Inspection of the simulations, however, reveals maximal spectral extent comparable with that dispayed in Fig.~\ref{fig1}(a).

\begin{figure}
\includegraphics[width=\columnwidth]{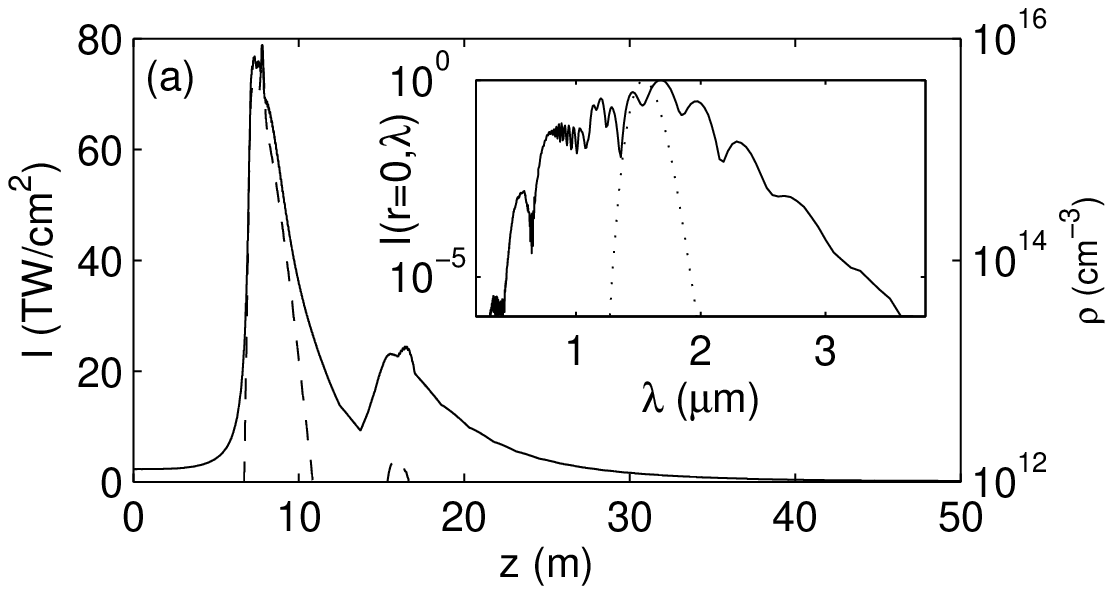}
\includegraphics[width=\columnwidth]{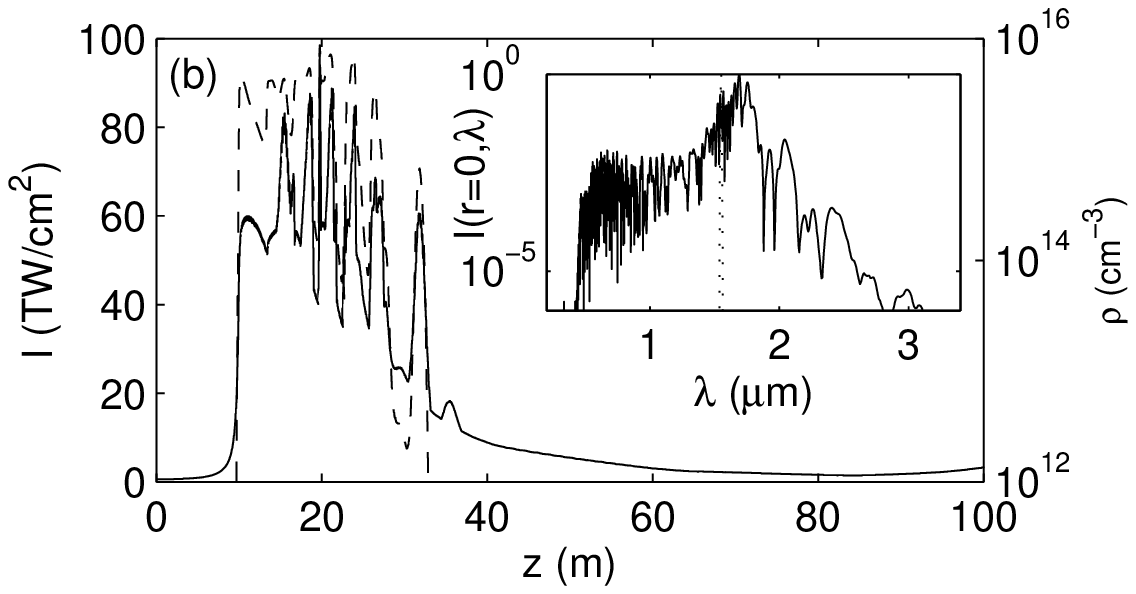}
\caption{\label{fig5} Peak intensities (solid curves, left-hand side scales) and peak electron densities (dashed curves, right-hand side scales) of 4~mm waisted pulses with different durations and ratios $P_{\rm in}/P_{\rm cr}$ at $\lambda_0=1550$~nm. The insets show on-axis spectra: dotted curves represent the initial spectra; the solid curves the spectra at the propagation distance $z_{\rm max}$ where maximal broadening is observed: (a) $t_p=20$~fs, $P_{\rm in}=15\times P_{\rm cr}$, $z_{\rm max}=20$~m; (b) $t_p=500$~fs, $P_{\rm in}=4 \times P_{\rm cr}$, $z_{\rm max}=40$~m.}
\end{figure}

At 1550~nm the influence of $t_p$ on the maximal intensity is much less pronounced, since we have $K=15$
[see Figs.~\ref{fig2}(b) and \ref{fig5}].
For all pulse durations, we indeed observe $I_{\rm max}\sim80$~TW/cm$^2$.
This can explain why the maximal spectral extension $\Delta \lambda_{\rm SC}/\lambda_0$ is always found between 1.5 and 2, regardless $t_p$ may be. The major difference lies in the filamentation range, which increases with the initial pulse duration. So, there is no significant change in SC generation between short and long pulses over large enough propagation scales.

\begin{figure}
\includegraphics[width=\columnwidth]{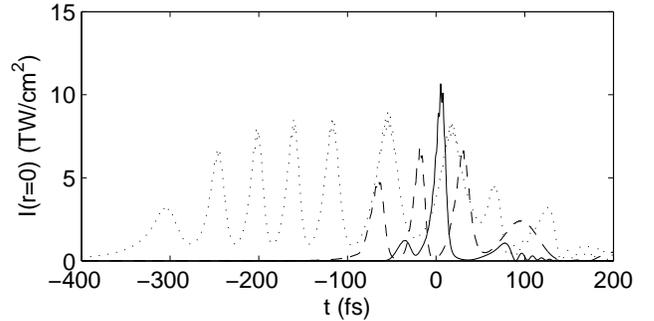}
\caption{\label{fig5bis} On-axis temporal profiles of pulses with different duration at $z_{\rm max}$: $t_p=20$~fs [solid line, parameters used in Fig.~\ref{fig5}(a)]; $t_p=127$~fs [dashed line, parameters used in Fig.~\ref{fig2}(b)]; $t_p=500$~fs [dotted line, parameters used in Fig.~\ref{fig5}(b)].}
\end{figure}

This last observation invites us to look at the temporal profiles upon propagation. If the spectral extent is comparable, we might also find similar temporal patterns. Indeed, the on-axis temporal profiles shown in Fig.~\ref{fig5bis} all exhibit structures with duration of 10-15~fs. It seems that the initial pulse length $t_p$ just determines how many of these peaks appear, or, in other words, how many focusing/defocusing cycles the pulse is able to develop upon propagation. Another indication for the change in the effective pulse duration upon propagation is provided by the curve of the intensity maximum in Fig.~\ref{fig5}(b): The first focusing cycle is halted at slightly lower intensities $\sim$ 60~TW/cm$^2$, because in this early stage the peak duration remains of the order of $t_p=500$~fs. At later stages, $I_{\rm max}$ increases as the pulse undergoes temporal compression.

\section{Short-range Propagation}

\label{Short-range}

\begin{figure}
\includegraphics[width=\columnwidth]{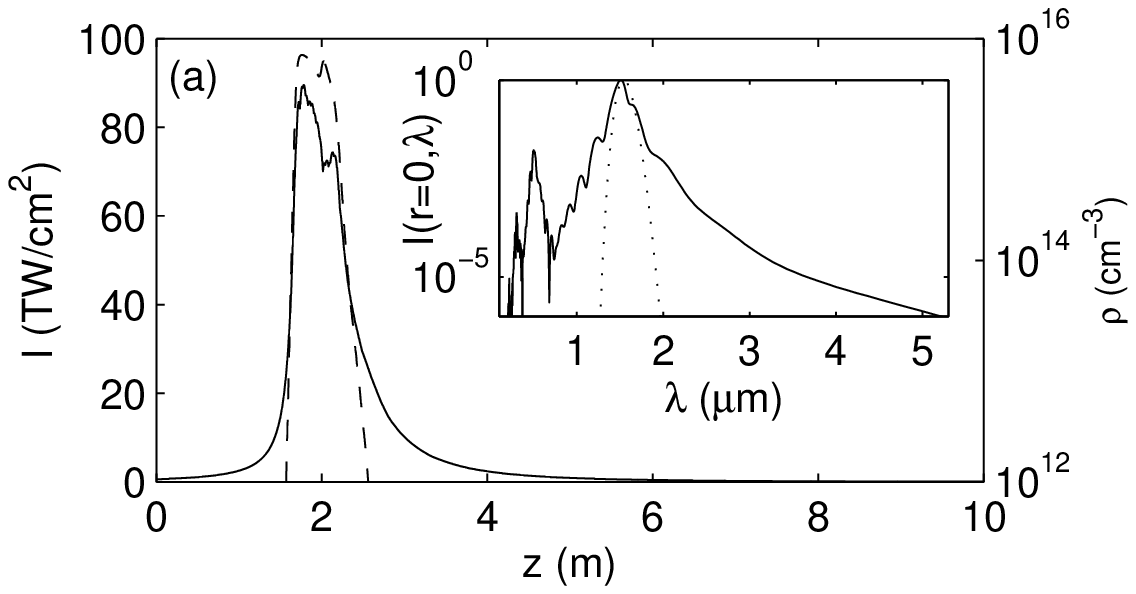}
\includegraphics[width=\columnwidth]{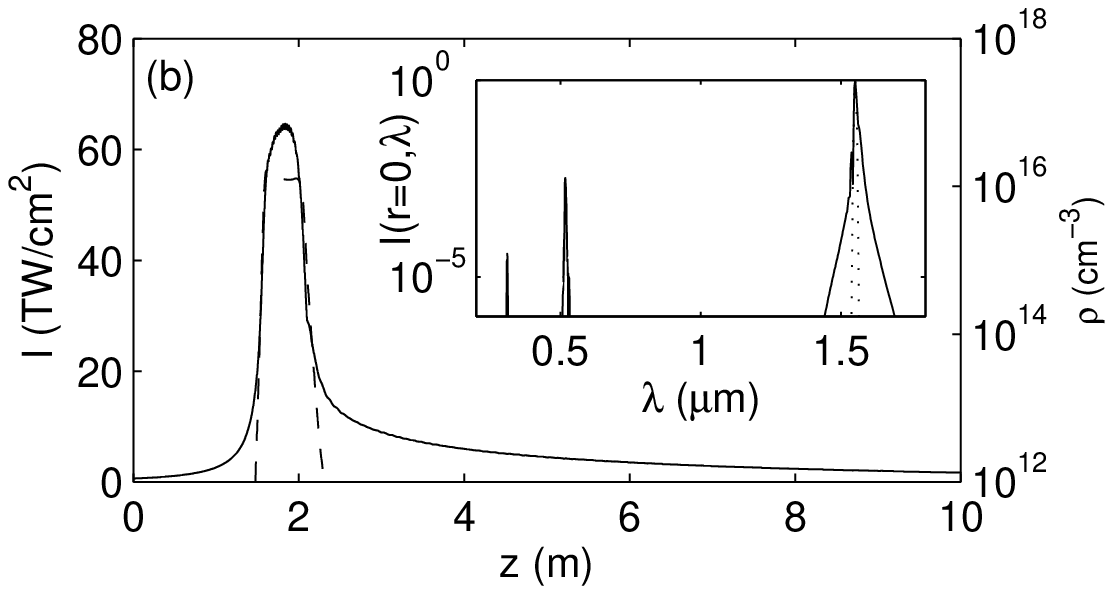}
\includegraphics[width=\columnwidth]{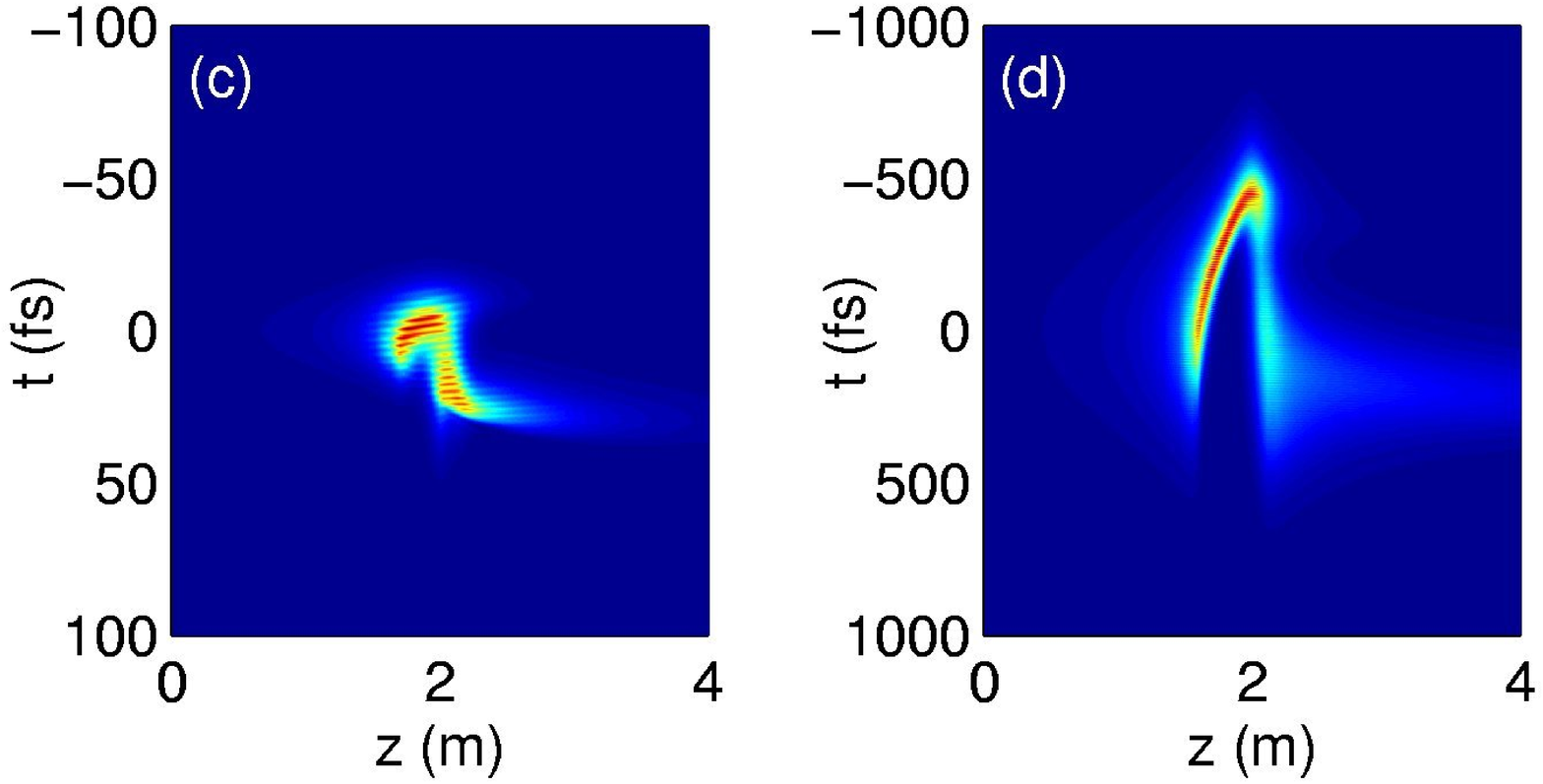}
\caption{\label{fig7} Peak intensities (solid curves, left-hand side scales) and peak electron densities (dashed curves, right-hand side scales) of 4~mm waisted pulses with different duration and ratio of input power over critical equal to 4 at $\lambda_0=1550$~nm propagating in focused geometry ($f = 2$ m).
The insets show on-axis spectra: dotted curves represent the initial spectra;
the solid curves the spectra at the distance $z_{\rm max}=2$~m
where maximal broadening is observed: (a) $t_p=20$~fs; (b) $t_p=500$~fs.
(c) and (d) show the respective temporal evolution for the pulses of (a) and (b) in the $(t,z)$ plane.}
\end{figure}

So far, we have analyzed free propagation dynamics where long filaments achieve temporal gradients and SC extents similar to those produced by initially much shorter pulses. Now, we force all pulses to cover the same short filamentation range through a focusing optics ($f=2$~m). Results are shown in Fig.~\ref{fig7} for $t_p = 20$ and 500~fs at 1550~nm. We can check that the maximum intensity $I_{\rm max}$ follows the theoretical expectations (\ref{estimate13}) involving $t_p$. The short filamentation range prevents the occurrence of several focusing/defocusing cycles, especially for the 500-fs pulse. Hence, although the pulse self-focuses, its temporal extent remains comparable with $t_p$, as we can see in Figs.~\ref{fig7}(c) and (d). The 20-fs pulse shows significant spectral broadening with a visible, broadened TH peak. The 500-fs pulse is
spectrally too narrow to generate supercontinuum over roughly 1-m filamentation range. Therefore, the fundamental and harmonic peaks clearly stay separated (note the occurrence of the fifth harmonics at 310 nm, which is self-consistently described by the UPPE model). Thus, the initial pulse duration strongly influences spectral broadening in configurations of short filamentation range mainly, which is consistent with the numerical results of Ref. \cite{Akozbek:njp:8:177}.

\section{Conclusion}

\label{conclusion}

In summary, we have revisited recent works on SC generation versus 
third harmonic emission, by showing from a complete UPPE model that spectral enlargements of femtosecond pulses in self-guiding regime are 
mostly driven by space-time focusing and self-steepening. TH 
generation, although changing the pump dynamics, affects the spectra 
to a limited extent only. This conclusion corrects some previous statements \cite{Mejean:apb:82:341}, based on a propagation model in which 
temporal steepening terms were analyzed separately from an envelope 
description for TH generation. Going one step beyond, we have demonstrated the important role of the saturation intensity in the frequency variations enlarging both TH and pump broadbands. The input pulse duration becomes a significant player in the spectral extents as long as pulses do not propagate 
too far, i.e., they do not let the temporal profiles of the pulse 
fluctuate so much that many peaks and sharp gradients can develop along the optical path. This property is mainly enlightened in short-range focused geometry and lost in long-range parallel geometry. Finally, the role of the central wavelength is preeminent: Our numerical simulations displayed evidence that SC clearly augments with the laser wavelength.

\end{document}